\documentstyle[11pt,newpasp,twoside,epsf]{article} 
\def\edcomment#1{\iffalse\marginpar{\raggedright\sl#1\/}\else\relax\fi}
\reversemarginpar

\begin{document} 

\title{Mm/submm observations of symbiotic binary stars} 

 \author{J. Miko{\l}ajewska$^1$, R.\,J. Ivison$^2$ \& A. Omont$^3$}

\affil{$^1$\, N. Copernicus Astronomical Center, Warsaw, Poland\\
$^2$\,Astronomy Technology Centre, Royal Observatory, Edinburgh, UK\\
$^3$\,Institut d'Astrophysique CNRS, Paris, France}

\begin{abstract} We present and discuss mm/submm observations of
quiescent \mbox{S-type} symbiotic systems, and compare them with
popular models proposed to account for their radio emission. We find
that the M giant mass-loss rates derived from our observations are
systematically higher than those reported for single M giants.
\end{abstract}

\section{Introduction}
To date $\sim$25\% of all symbiotic stars have been detected in the
cm-wave radio band. In practically all cases, the radio emission is
consistent with {\it ff} radiation from ionised gas (Seaquist, \&
Taylor 1990). Seaquist, Taylor, \& Button (1984) proposed a simple
binary model (the STB model) in which the radio emission originates
from the red giant wind, partly ionized by the hot companion. The
predicted radio spectrum turns over from optically thick to optically
thin emission at a frequency, $\nu_{\rm t}$, which is related to the
binary parameters. The observations of $\nu_{\rm t}$ in quiescent
\mbox{S-type} systems with known orbital parameters thus provides a
critical test of the STB model. Unfortunately, the spectral turnovers
have been thusfar determined only in either D-type systems (with
unknown orbital periods) or \mbox{S-type} systems recovering from nova
outbursts, e.g.\ AG~Peg (e.g.\ Ivison, Hughes, \& Bode 1995).

In the following, we present and discuss results of mm/submm
observations of a sample of \mbox{S-type} symbiotic stars collected at
1.3\,mm with the IRAM 30-m MRT (February 1997), and at 2, 1.3, 0.85 and
0.45\,mm with the SCUBA submm camera on the JCMT (1997--98).

\begin{figure}[t] 
\plotone{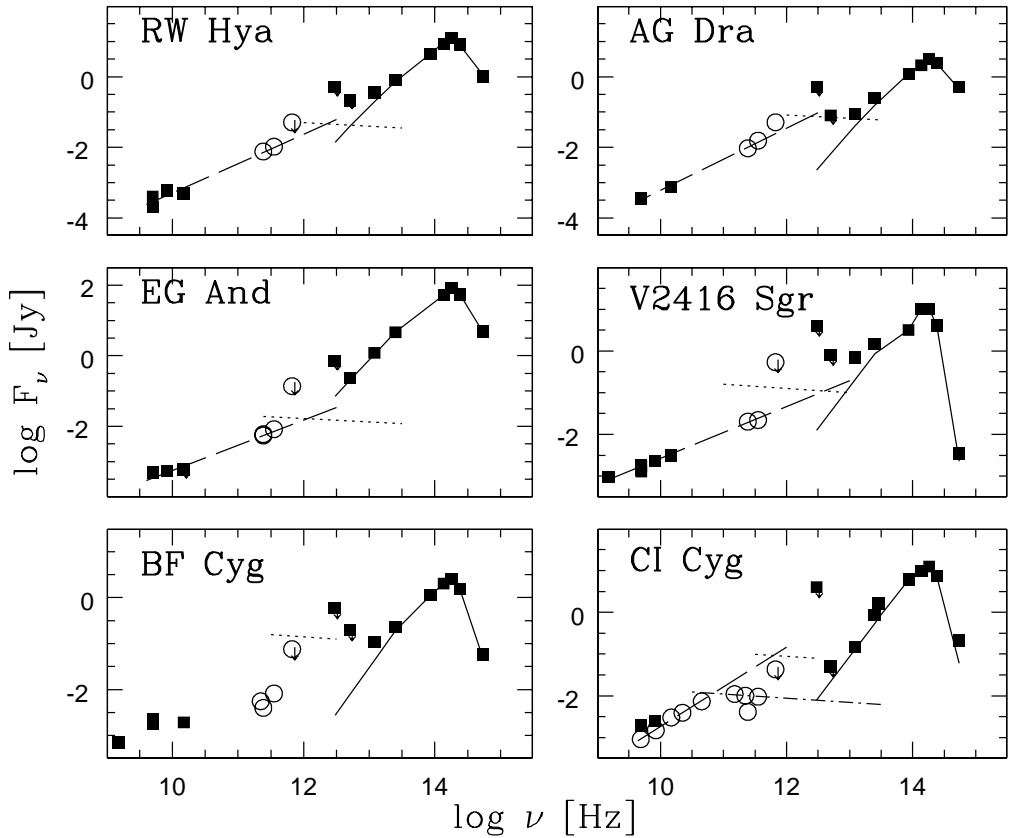}
\caption{Continuum spectra of quiescent \mbox{S-type} systems. Circles
refer to our new data, squares to old data. The spectra are split into
optically thick/thin {\it ff} emission (dashed, dotted and dot-dashed
curves, respectively; see text) and the giant photosphere (solid
curve).}\end{figure}

\section{Testing the STB model}

We have selected 20 \mbox{S-type} systems, all quiescent at the time
of our observations. Most of these systems have been studied
intensively in the optical and UV; binary periods are known for 14 of
them, and for 7 systems we also have spectroscopic orbits. For more
details we refer to Miko{\l}ajewska, \& Ivison (2001, hereafter MI01),
and Miko{\l}ajewska, Ivison, \& Omont (2002, hereafter MIO).

Comparison of 1.3-mm flux densities with published H$\beta$ fluxes
(MIO) indicates that the radio emission from quiescent \mbox{S-type}
systems is optically thick at least up to 1.3\,mm, in agreement with
the STB model which predicts a spectral turnover at submm wavelengths
for such systems (Seaquist, Krogulec \& Taylor 1993, hereafter
SKT93). Our mm/submm continuum spectra (Fig.~1) are also consistent
with optically thick thermal emission for all systems except
CI~Cyg. To constrain $\nu_{\rm t}$ and estimate the binary separation
for these systems, the optically thin {\it ff} emission in the mm
range (dotted curves in Fig.~1) was estimated using the H\,{\sc i}
{\it ff+bf} emission measure derived from optical and UV data.  The
resulting binary separations are in good agreement with the values
derived from known spectroscopic orbits (see Fig.~4 of MIO). The
mm-wave emission also shows some correlation with the mid-IR flux, and
the radio luminosity increases with the $K-[12]$ colour, which
indicates that both the ionised gas and warm dust are involved in the
mass flow, and suggests that the cool giant may be the source of this
material.  We note, however, that the consistency between the
mm/submm-wave data and the STB binary model revealed by the work of
MIO is not based on direct measurements of the spectral turnover, only
on an optically thin emission measure inferred from optical/UV
spectroscopy which, in addition, was not collected simultaneously with
the radio observations.

Recently, MI01 have determined for the first time the 0.85\,mm--6\,cm
spectral energy distribution (SED) of a prototypical \mbox{S-type}
symbiotic system, CI~Cyg, during quiescence (Fig.~1). Unfortunately,
comparison of the binary separation and the Lyman continuum photon
luminosity derived from this SED with the known orbital and stellar
parameters of CI~Cyg rules out both the STB model as well as models
based on the interaction of winds from the binary companions. In
particular, the {\it ff} turnover frequency determined from the SED in
a model-independent manner (Fig.~1) over-estimates the binary
separation by a factor of $\sim$36, while the optically thin {\it ff}
emission measure under-estimates $L_{\rm ph}$ by a factor of $\sim$20
relative to the value inferred from optical/UV spectroscopy
(Fig.~1). One possible cause of this low, optically thin radio
emission is that CI~Cyg is one of the few symbiotic systems in which
the M giant shows strong tidal distortion, and loses mass via
Roche-lobe overflow rather than via a stellar wind. In such a case
the bulk of the Balmer H\,{\sc i} emission is most likely formed in
dense material in the orbital plane (stream and/or accretion disc)
whereas the radio emission and high-excitation forbidden lines arise
from lower density regions in polar directions.

If the giant does fill its Roche lobe, the M giant's wind in
\mbox{S-type} systems is likely to be focused towards the secondary
and/or towards the orbital plane. In particular, gravitational
interaction of the cool giant's wind with the secondary can produce an
equatorial-to-polar density contrast as large as 100--1000, giving
rise to a bipolar geometry of the symbiotic nebulae (Gawryszczak,
Miko{\l}ajewska, \& R{\'o}{\.z}yczka 2002a). Such a geometry results
in significantly different shapes of the H\,{\sc ii} region, and the
emerging SEDs, compared with their STB counterparts (Gawryszczak,
Miko{\l}ajewska, \& R{\'o}{\.z}yczka 2002b). Moreover, in
high-inclination systems with relatively low $L_{\rm ph}$ and high
$\dot{M}/v$ ($X \sim 1$ in the STB model) two radio lobes are
expected. We note here that the recently resolved radio emission from
AG~Dra cannot be accounted for by the STB model (Miko{\l}ajewska 2002,
and references therein) although MIO did not reveal any inconsistency
between its SED (in particular, $\nu_{\rm t}$, Fig.~1) and the STB
model.

\section{Mass-loss rates}

\begin{figure}\plotone{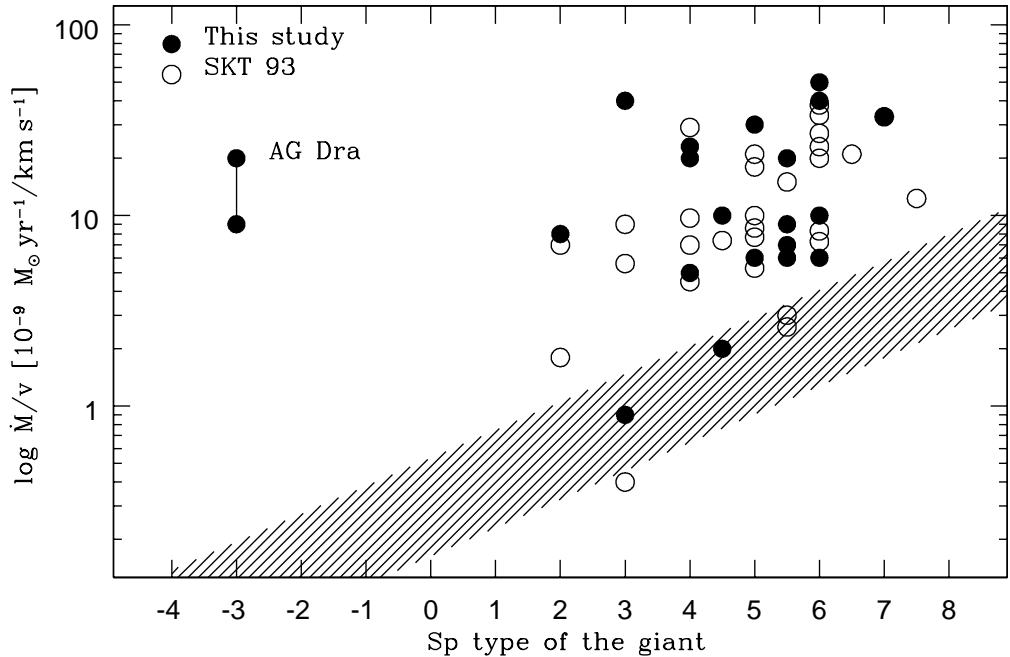} 
\caption{Comparison of $\dot{M}/v$ estimates for symbiotic binaries
with those for single field giants (shaded area).}
\end{figure}

The M giant mass-loss rates derived from 1.3-mm flux densities,
applying the WB relation (Wright, \& Barlow 1975), are systematically
higher than those reported for single M giants (MIO; Fig.~2). Similar
results were obtained by SKT93 based on cm-wave observations and
Kenyon, Fernandez-Castro, \& Stencel (1988) from analysis of {\em
IRAS} data. MIO also found a correlation between the giant mass-loss
rate and the hot component's luminosity, $L_{\rm h}$, which may
suggest that illumination of the outer atmosphere of the giant is an
important effect. The WB relation under-estimates $\dot{M}$ if the wind
is only partially ionized, which seems to be the case for most if not
all quiescent symbiotic systems. SKT93 estimated the magnitude of this
under-estimate based on the STB model as a factor of 2, 1.5 and 1.15
for $X$ = 0.5, 1, and 5, respectively. In fact, the mass-loss rates
can be under-estimated by a factor of 10 or more if the ionized region
is highly asymmetric (Gawryszczak et al.\ 2002b). Since this effect
increases with decreasing $L_{\rm ph} \propto L_{\rm h}$, it can also
account for the correlation between the mass loss estimates based on
spherically symmetric models (such as WB and STB) and $L_{\rm h}$.

\acknowledgments This research was partly founded by KBN Research
Grant No.  5\,P03D\,019\,20, and by the JUMELAGE program `Astronomie
France-Pologne' of CNRS/PAN.

\end{document}